\documentclass{jaa}
\usepackage{natbib}
\usepackage{xcolor}
\definecolor{xlinkcolor}{cmyk}{1,1,0,0}

 \usepackage[bookmarks=false,         
     pdfnewwindow=true,      
     colorlinks=true,    
     linkcolor=xlinkcolor,     
     citecolor=xlinkcolor,     
     filecolor=xlinkcolor,  
     urlcolor=xlinkcolor,      
final=true 
 ]{hyperref}

\usepackage{graphicx}
\usepackage{lastpage}
\usepackage{array}

\newcommand{\degr}{\ensuremath{^\circ}}
\newcommand{\asat}{{\em AstroSat}}
\newcommand{\cyg}{Cyg~X--1}
\newcommand{\sw}[1]{\texttt{#1}}

\newcommand{\hocc}{\ensuremath{h_\mathrm{occ}}}
\newcommand{\thw}{\ensuremath{t_\mathrm{hw}}}
\newcommand{\tbin}{\ensuremath{t_\mathrm{bin}}}

\newcolumntype{x}[1]{%
>{\centering\hspace{0pt}}p{#1}}%

\hypersetup{draft}
\begin{document}\sloppy

\title{Using collimated CZTI as all sky X-ray detector based on Earth Occultation Technique }


\author{\href{https://orcid.org/0000-0003-1275-1904}{Akshat Singhal}\textsuperscript{1,*},
\href{https://orcid.org/0000-0002-7176-6690}{Rahul Srinivasan}\textsuperscript{1,2,6,7,8}, 
\href{https://orcid.org/0000-0002-6112-7609}{Varun Bhalerao}\textsuperscript{1},
\href{https://orcid.org/0000-0003-3352-3142}{Dipankar Bhattacharya}\textsuperscript{3},
\href{https://orcid.org/0000-0003-0833-0533}{A. R. Rao}\textsuperscript{4},
\and \href{https://orcid.org/0000-0002-2050-0913}{Santosh Vadawale}\textsuperscript{5}
}
\affilOne{\textsuperscript{1} Indian Institute of Technology Bombay, Mumbai, India\\}
\affilTwo{\textsuperscript{2} Universit\'e C\^ote d'Azur, Nice, France\\}
\affilFive{\textsuperscript{5} Observatoire de la C\^ote d'Azur, Nice, France\\}
\affilSix{\textsuperscript{6} CNRS, Nice, France\\}
\affilSeven{\textsuperscript{7} Laboratoire Lagrange, Nice, France\\}
\affilEight{\textsuperscript{8} Laboratoire Art\'emis, Nice, France\\}

\affilThree{\textsuperscript{3} The Inter-University Centre for Astronomy and Astrophysics, Pune, India\\}
\affilFour{\textsuperscript{4} Tata Institute of Fundamental Research, Mumbai, India\\}
\affilFive{\textsuperscript{5} Physical Research Laboratory, Ahmedabad, Gujarat, India\\}


\twocolumn[{

\maketitle

\corres{akshats@iitb.ac.in}

\msinfo{1 January 2015}{1 January 2015}

\begin{abstract}
All-sky monitors can measure the fluxes of astrophysical sources by measuring the changes in observed counts as the source is occulted by the Earth. Such measurements have typically been carried out by all-sky monitors like {\em CGRO}-BATSE and {\em Fermi}-GBM. We demonstrate for the first time the application of this technique to measure fluxes of sources using a collimated instrument: the Cadmium Zinc Telluride detector on \asat. Reliable flux measurements are obtained for the Crab nebula and pulsar, and for \cyg\ by carefully selecting the best occultation data sets. We demonstrate that CZTI can obtain such measurements for hard sources with intensities $\gtrsim1$Crab. 
\end{abstract}


}] 


\doinum{12.3456/s78910-011-012-3}
\artcitid{\#\#\#\#}
\volnum{000}
\year{0000}
\pgrange{1--}
\setcounter{page}{1}
\lp{\pageref{LastPage}}

\section{Introduction}
 
Modern high energy instruments use a variety of techniques including collimators, coded aperture masks, and grazing incidence / multi-layer optics to characterise astrophysical sources and measure their position, flux, and spectra. Such instruments face an inherent trade-off between sky coverage and localisation accuracy. All-sky monitors like {\em CGRO}-BATSE and {\em Fermi}-GBM can characterise transient sources based on relative flux intensities on various detectors. But their methods rely on the fact that the effect of all other sources in the sky can be corrected for by using pre- and post-transient data. Such open-detector all-sky monitors have very limited abilities for studying persistent astrophysical sources.

One such all-sky monitor is the Cadmium Zinc Telluride Imager \cite[CZTI][]{bhalerao2017cadmium} on board \asat\ \citep{singh2014astrosat}. CZTI is a coded aperture mask instrument sensitive in the 20--200~keV band. At energies above $\sim100$~keV, the instrument and satellite structures become transparent to radiation, giving CZTI sensitivity to sources well outside the primary field of view. CZTI data have been used to detect over 400 Gamma Ray Bursts (GRBs) in the 5 years since launch 
\citep{sharma2020search}, and even study the Crab pulsar at angles from 5\degr\ to 70\degr\ from the principal axis 
\citep{2021arXiv210108650A}. However, CZTI faces the same limitations as other all-sky monitors for studying persistent astrophysical sources.

A workaround for studying such sources which lack rapid intrinsic variability is to leverage any extrinsic variations in their flux. A popular method for this is to observe sources as they are occulted by the Earth or the Moon. The transition of the source behind the Earth's limb (ingress) causes a rapid step-like decrease in the overall counts measured by the detector, and likewise the transition from behind the limb to visibility (egress) causes a rapid increase in the counts. Lunar occultation  techniques have even been used in ground-based radio telescopes with great success \citep[see for instance][]{1973AJ.....78..673K}. Earth occultation studies have been successfully leveraged for studying high-energy sources by space-based detectors like {\em CGRO}-BATSE~\citep{harmon2002burst} and {\em Fermi}-GBM~\citep{wilson2012three}. In this work, we demonstrate the application of the Earth Occultation Technique (EOT) for measuring source fluxes using data from the CZTI.

\section{Overview of the Earth Occultation Technique}
We begin with a brief overview of the formulations of EOT. For more details, we refer the readers to \citet{harmon2002burst} and \citet{zhang2013searching}. 

\begin{figure*}[htb]
\centering
\includegraphics[width = 0.6\linewidth]{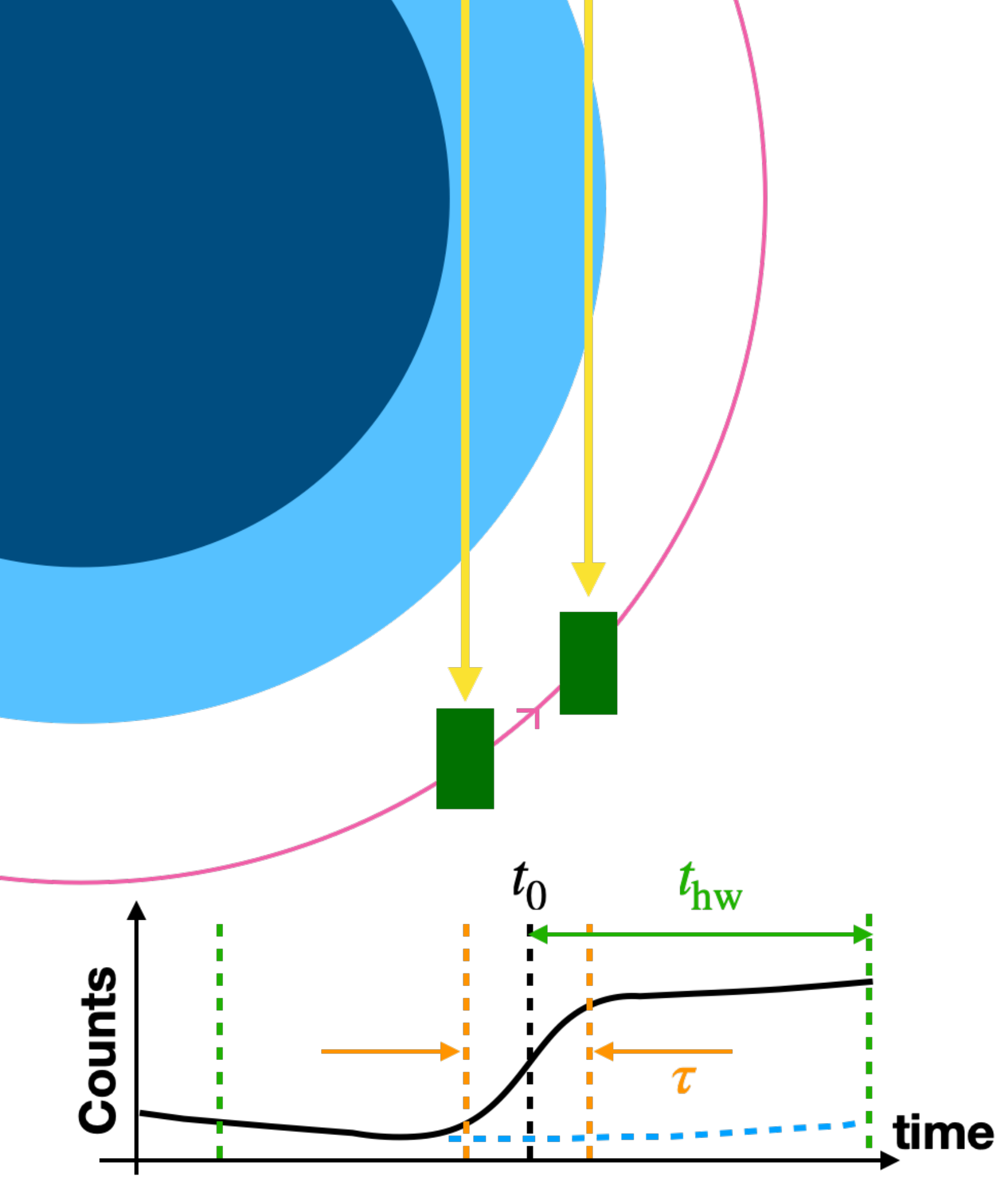}
\caption{A schematic of a source egress and the corresponding observed light curve. The top part shows the Earth (dark blue circle) and its atmosphere (slight blue annulus). \asat\ (green rectangle) is orbiting the Earth in a counterclockwise sense, as shown by the violet curve. Vertical yellow arrows denote the parallel rays of light coming from a distant astrophysical source. The bottom part shows the corresponding light curve (solid black line) observed by CZTI, which we model as $M(t-t_0)$ (Equation~\ref{Eq:TotalFit}). The dashed blue line marks the background-only counts curve ($b_g$) expected if there was no source egress. Vertical dashed lines mark various times of interest: the dashed black line nominal occultation time $t_0$ when the source is at a height \hocc\ above the Earth's limb, dashed orange lines show the window of duration $\tau$ when the source flux rises from 10\% to 90\% level, and dashed green lines show the complete window of interest used in our fitting procedures.}
\label{fig:occult}
\end{figure*}

For a satellite in say a 650~km Low Earth Orbit (LEO), the Earth subtends an angle of $\sim130\degr$ on the sky. The inclination and precession of the satellite orbit determine which astrophysical sources can get occulted by the Earth: for instance, for an equatorial orbit, objects near the celestial poles are never obscured by the Earth. For the 6\degr\ orbital inclination of \asat, sources with declination up to approximately $\pm70\degr$ can get occulted at some point, while objects with declination in the range of about $\pm60\degr$ will get occulted in each orbit. In each orbit around the Earth, the satellite can observe one ingress and one egress of these sources (Figure~\ref{fig:occult}).

For very bright objects, occultations can be used to measure flux, spectra, and to measure the location of the source in the direction perpendicular to the Earth's limb. In this work, we assume that the source position and approximate spectral shape are known, and attempt to measure the source flux (spectral normalisation).

\subsection{Time and duration of occultation}\label{sec:ToE} 
Using the known source coordinates and orbital parameters of the satellite, we can calculate the time of occultation ($t_0$) of each ingress and egress of the source. At high altitudes ($\gtrsim 100$~km) the Earth's atmosphere is largely transparent to X-rays, and becomes increasingly opaque closer to the surface. This has two consequences for occultation studies. First, the nominal ingress and egress do not occur at the geometric limb of Earth, but some distance further out. Second, the decrease / increase in source flux are not sharp step functions, but more gradual transitions. To incorporate these effects in our work, we define an effective height \hocc\ at which the source flux decreases to 50\% of its unobstructed value. For the 20--200~keV energy range of CZTI, we adopt \hocc = 70~km. 

It is to be noted that shape of the earth and energy dependent \hocc\ can affect the calculations. However we used a simpler approach of spherical earth with constant \hocc\ for its simplicity. In comparison to alternate approach of adding more degrees of freedom or removing the data near the transition resulted in more errors. 

\begin{figure}[ht]
\centering
\includegraphics[trim=0 50 0 50, clip,width = 0.8\linewidth]{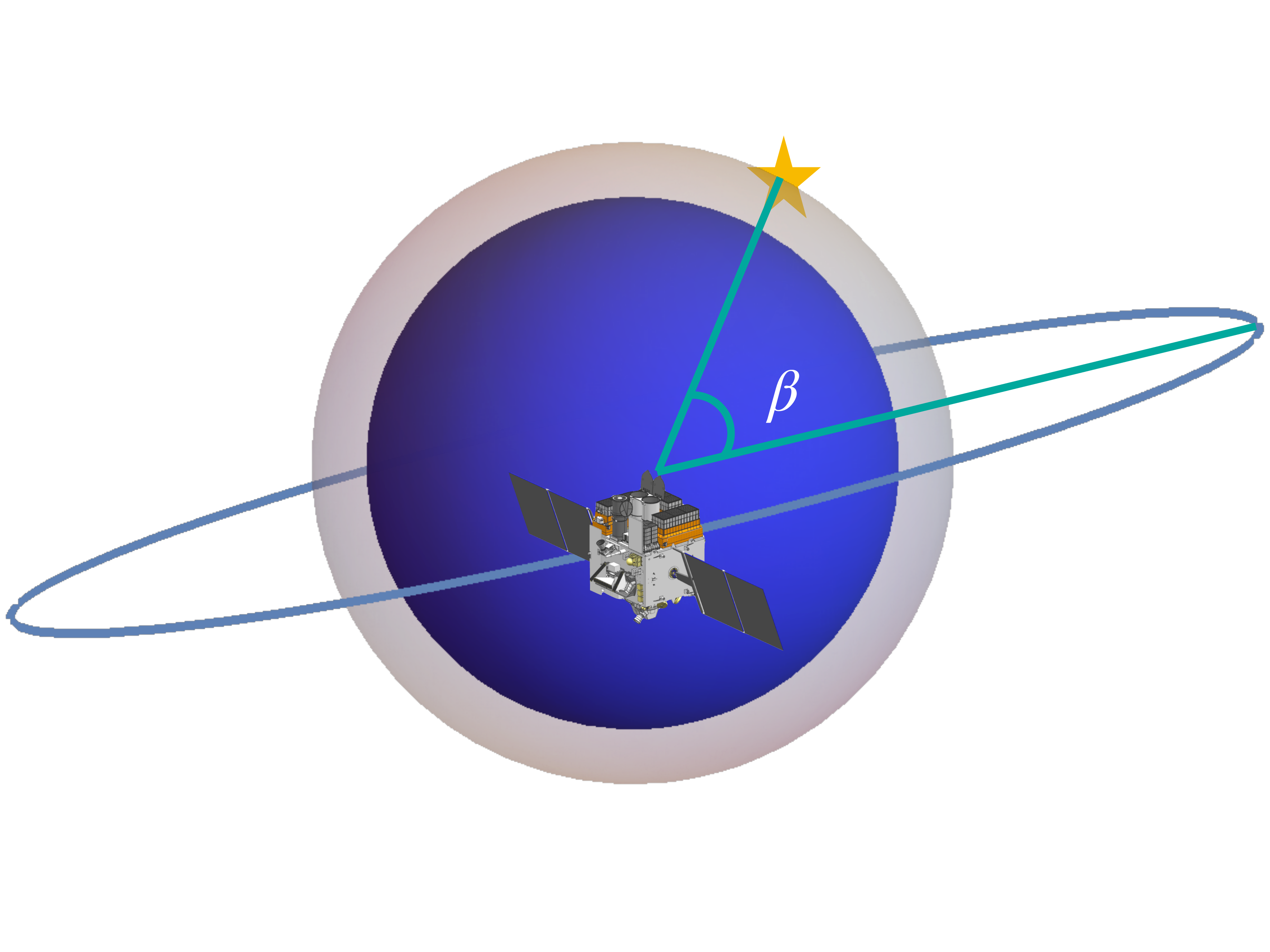}
\caption{Schematic diagram for relative orbital declination. This diagram will be replaced by our own. This is from Fermi.}
\label{fig:beta}
\end{figure}

To account for the gradual transition, we define a characteristic timescale $\tau$ over which the source flux in an ingress drops from 90\% to 10\% of its unobstructed value. The orbital inclination of \asat\ and the declination of the source also play a role in determining the transition timescale. Due to projection effects, sources further away from the orbital plane of the satellite appears to pass through oblique layers of the atmosphere, extending the duration of the transition. This is accounted for by defining an ``orbital latitude'' $\beta$ \citep{harmon2002burst}, the projected angle between orbital plane and the line joining the source to the centre of the Earth (Figure~\ref{fig:beta}). The transition timescale is then given by $\tau = \tau_0 \sec \beta$, where $\tau_0$ is the minimum transition timescale, observed for sources in the orbital plane.

\subsection{Background and observed flux}\label{sec:bag} 
The background for CZTI comprises of various components including cosmic X-ray background, Earth albedo, charged particle induced background, and electronic noise. In addition to these, for the purpose of our EOT analysis, photons from the on-axis source also contribute to background. The background level is not constant: it varies slowly through the orbit (dominated by proximity to the South Atlantic Anomaly), and also shows variations across orbits (dominated by orbital precession and solar activity). It is known that these background variations for CZTI occur on timescales of several hundreds to thousands of seconds, and is well-approximated by a quadratic function \citep{sharma2020search}\citep{anumarlapudi2020prompt}. Hence we too model background as a quadratic function, $b_g(t) = a(t-t_0)^2 + b(t-t_0) + c$.

Various models have been used for the smooth source intensity transition. We adopt a simplistic error function, such that the transition light curve $T_E$ for an egress is given by,

\begin{align}
T_E(t-t_0)   &= h\left(\frac{1}{2}+\frac{1}{2}\rm{erf\left(\frac{1.812(t-t_0)}{\tau}\right)}\right), \\
 &= \frac{h}{2\sqrt{\pi}}\int_{-\infty}^{\frac{1.812(t-t_0)}{\tau}}e^{-x^2}\,dx \label{err}
\end{align}
where erf is the error function, $\tau$ is the transition timescale discussed in \S\ref{sec:ToE}, $t_0$ is the nominal transition time when the source cross the altitude \hocc, and $h$ is the average count rate from the source being occulted. We note that $h$ is negative for source ingress. The net model $M$ for the occultation is thus given by,
\begin{equation}
    M(t-t_0) = b_g(t-t_0) + T_E(t-t_0)
    \label{Eq:TotalFit}
\end{equation}

Lastly, we note that we perform fits to observed data in the range $t_0 \pm \thw$, where \thw\ stands for the half-window time.

\section{Methodology}\label{sec:method}
\begin{figure}[ht]
    \centering
    \includegraphics[width=\columnwidth]{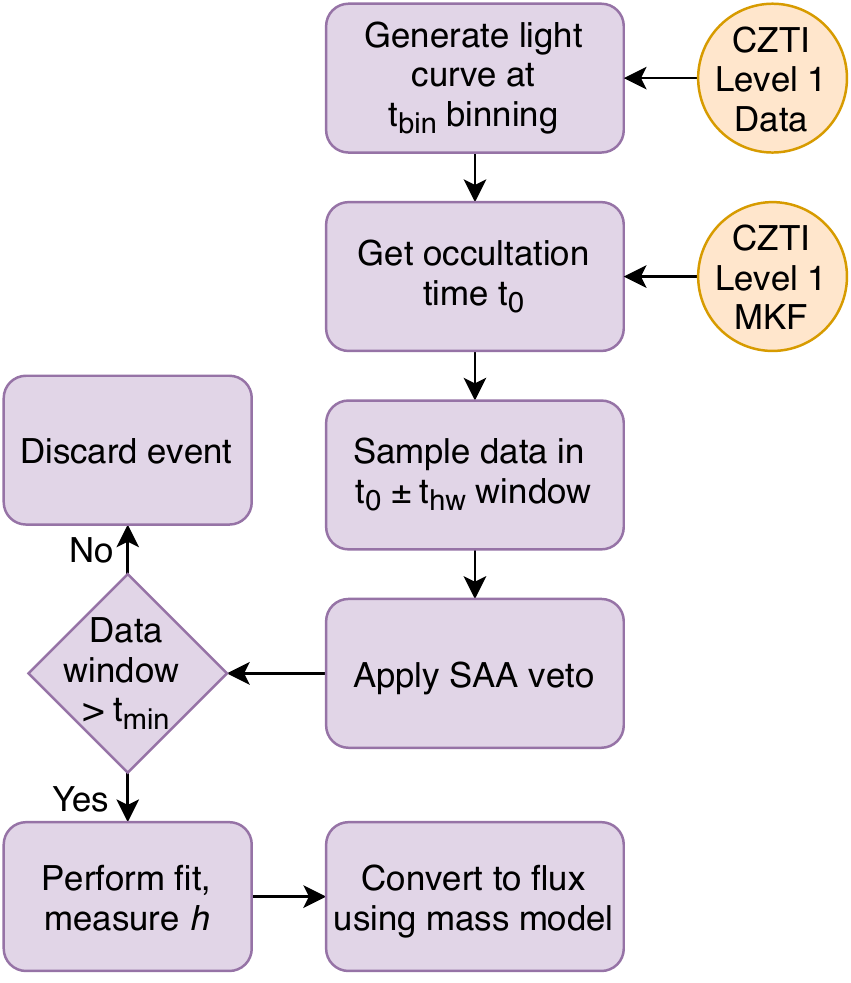}
    \caption{Flowchart for the Earth occultation pipeline for observing X-ray flux using \asat-CZTI data}\label{fig:flowchart-2}
\end{figure}

In this section, we discuss the specific details for EOT analysis of CZTI data. The overall process flow chart is shown in Figure~\ref{fig:flowchart-2}.

\subsection{Preparing the data}\label{sec:dataprep}
CZTI consists of four identical independent quadrants (labeled A, B, C, D), each containing sixteen detectors with 256 pixels each \citep{bhalerao2017cadmium}. All observed data are downloaded in event mode, noting the time, energy, and position of interaction for every incident photon. We reduce the Level 1 data using the default CZTI pipeline\footnote{CZTI pipeline: \url{http://astrosat-ssc.iucaa.in/?q=cztiData}}, with one tweak. By default, the pipeline discards data in intervals when the on-axis source is close to or occulted by the Earth. Since we are not interested in the analysis of on-axis sources, we disable this flag in the pipeline. We use \sw{cztbindata} to bin the data at \tbin\ seconds, creating livetime-corrected light curves complete with count rate uncertainties for each bin. We limit ourselves to using only ``grade 0'' pixels. 

We calculate the source occultation time $t_0$ based on the source coordinates and the location of the satellite with respect to the Earth. The latter is obtained from the CZTI Level 1 \sw{MKF} file. Our window of interest is $t_0 \pm \thw$ as discussed above. 

Like all high energy instruments, CZTI stops collecting data when the satellite is passing through the South Atlantic Anomaly (SAA). We identify the start and end of the SAA as points in the light curve where the count rate drops to and rises from zero respectively. Data obtained very close to SAA are often very noisy. Hence, we add another 10~s ``buffer'' time on either side of the SAA. If any part of our window of interest overlaps with the SAA passage or the buffer time, we discard the overlapping part. After this cut, we require that at least $t_\mathrm{min} = 0.417\thw$ seconds of data must be present on either side of $t_0$, else the entire event is rejected\footnote{The fraction 0.417 = 5/12 comes from backward compatibility during code development.}.

\subsection{Signal modelling}\label{sec:sigmod}
\begin{figure}[htb]
    \centering
    \includegraphics[width = 0.8\linewidth]{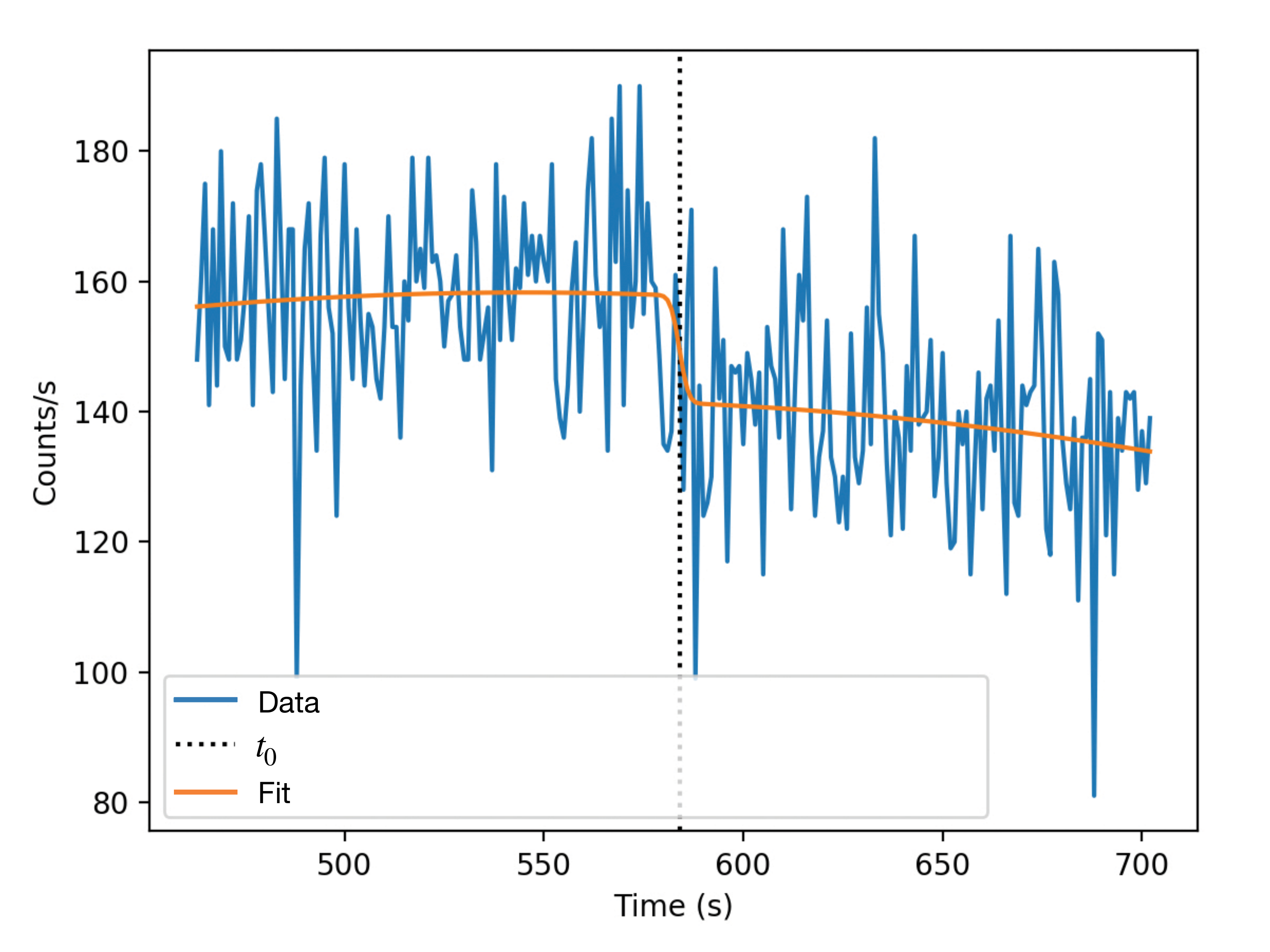}
    \caption{Fitting of Eq.~(\ref{Eq:TotalFit}) to measure the count rate from Crab in CZTI Quadrant A.}
    \label{fig:Fits}
\end{figure}

We now fit Equation~\ref{Eq:TotalFit} to the light curves in the $t_0 \pm \thw$ time range (Figure~\ref{fig:Fits}). The source model $M$ has six parameters: the source counts $h$, the occultation timescale $\tau$, the time of occultation $t_0$, and the three quadratic parameters $a$, $b$, $c$ for the background. Of these, we keep $\tau$ and $t_0$ fixed, and solve for the remaining four using the least-squares solver \sw{curve\textunderscore fit} implemented in the \sw{scipy.optimize} python package. 
A sample fit for the ingress of Crab is shown in Figure~\ref{fig:Fits}. The blue lines denote the light curve, while the solid orange line is the best-fit result. The vertical black dashed line marks $t_0$, the expected time of the occultation.

We find that these fits can be skewed by presence of outliers in the light curve, and outlier rejection is crucial for getting good fits. We implement this process iteratively: first, the model $M(t-t_0)$ is fit to the light curve in the $t_0 \pm \thw$ range. All data points are weighted by the uncertainties calculated by the CZTI pipeline. We select a threshold $k$ such that $k\sigma$ outliers in this fit are rejected. The uncertainties on individual data points are ignored in this step. The fitting procedure is repeated, and outliers from the new fit are rejected again. We repeat this procedure 10 times or until there are no more $k\sigma$ outliers. Experimenting with our data, we found that the best results are obtained when $k = 2$, corresponding to on average of 15\% of data rejected.


In addition to the uncertainty in fit parameters returned by the solver, we use another method to assess the significance of our fit. We exclude the central $t_0 \pm \tau$ region of our window of interest, and evaluate the residuals obtained by subtracting our best-fit model from the data. We calculate the standard deviation ($\sigma_\mathrm{counts}$) of these residuals for the pre-$t_0$ and post-$t_0$ regions. Each of these regions consists of $N_\mathrm{bin} = (\thw - \tau)/\tbin$ independent bins, so the uncertainty in estimating the background is approximately $\sigma_\mathrm{counts} / \sqrt{N_\mathrm{bin}}$. Since $h$ roughly corresponds to difference between the two background levels, the typical uncertainty of a detection of an occultation will be given by,
\begin{equation}
    h_\mathrm{min} \approx \frac{\sigma_\mathrm{counts} \sqrt{2}}{\sqrt{N_\mathrm{bin}}}
    \label{eq:hmin}
\end{equation}

\subsection{Selecting parameters}\label{sec:selpar}
The algorithm described above still leaves several free parameters to be selected: the energy range, \tbin, and \thw. In most of our work, we use the full CZTI data range from 20--200~keV, with the exception of specific test discussed in \S\ref{sec:highenergy}.

We explored various values for \tbin\ from 1~s to 10~s. We observed that as \tbin\ increases,  the flux posterior broadens leading to progressively imprecise measurements. One possible explanation is that since the transition period is $\mathcal{O}(10$~s), comparable values of \tbin\ smear out this transition period, and the fitting function --- which is simply evaluated at bin centres --- cannot accurately recover the change in count rates. As a result, we fix \tbin=1~s for our analysis. 

We tested various values of \thw\ from 100~s to 1000~s. We observe that if \thw\ is small, the estimation of background suffers, increasing $h_\mathrm{min}$ (Equation~\ref{eq:hmin}), equivalently increasing the uncertainty in $h$. On the other hand, if the window is too wide, the background can no longer be accurately modelled by a quadratic. We adopt \thw = 500~s for our analysis.

Based on these values, we find the typical value of $h_\mathrm{min}$ is around 0.9 counts for 20--200~keV energy range.

\begin{figure*}[htbp]
    \centering
    \includegraphics[width = 6cm]{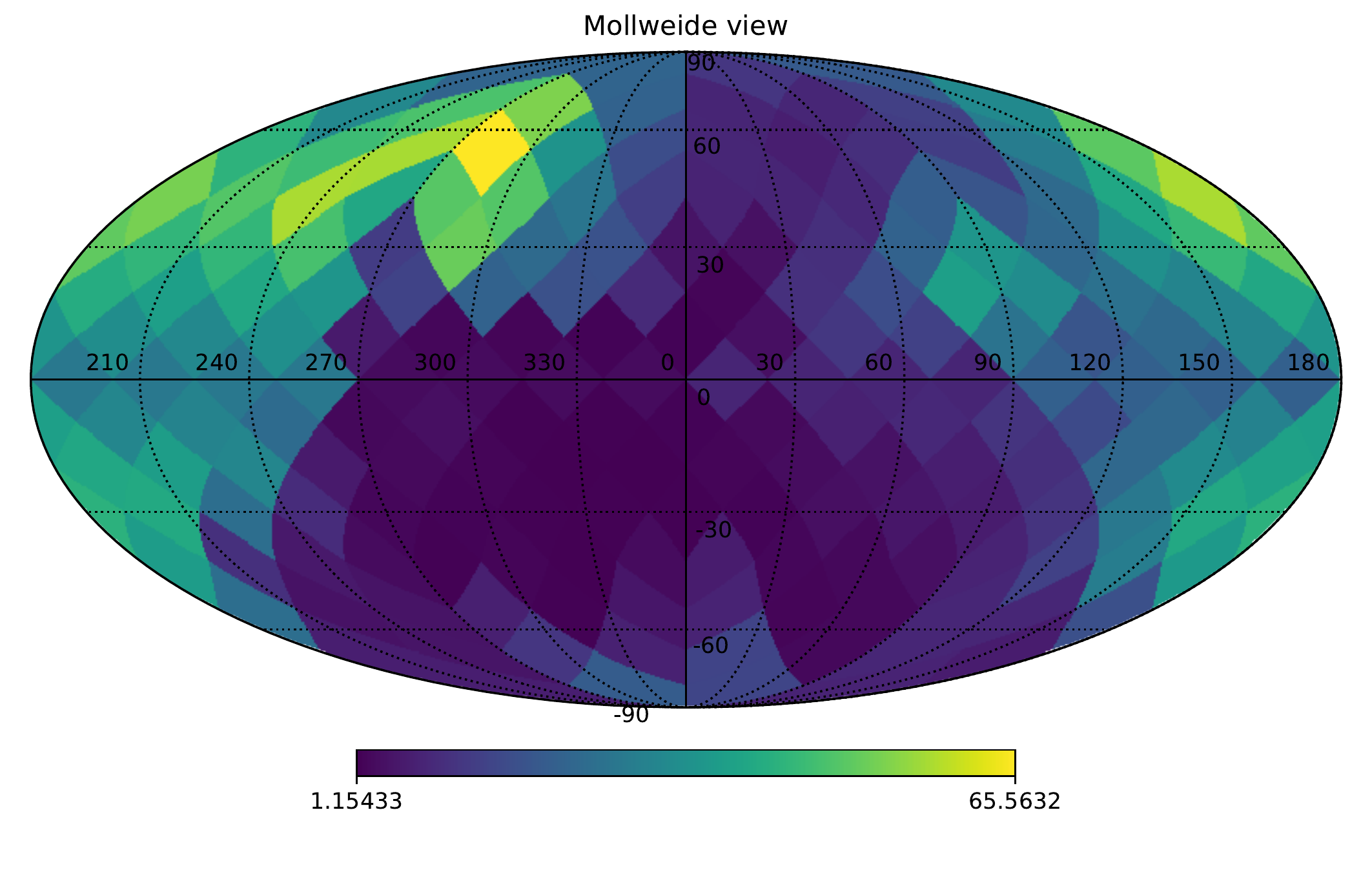} \includegraphics[width = 6cm]{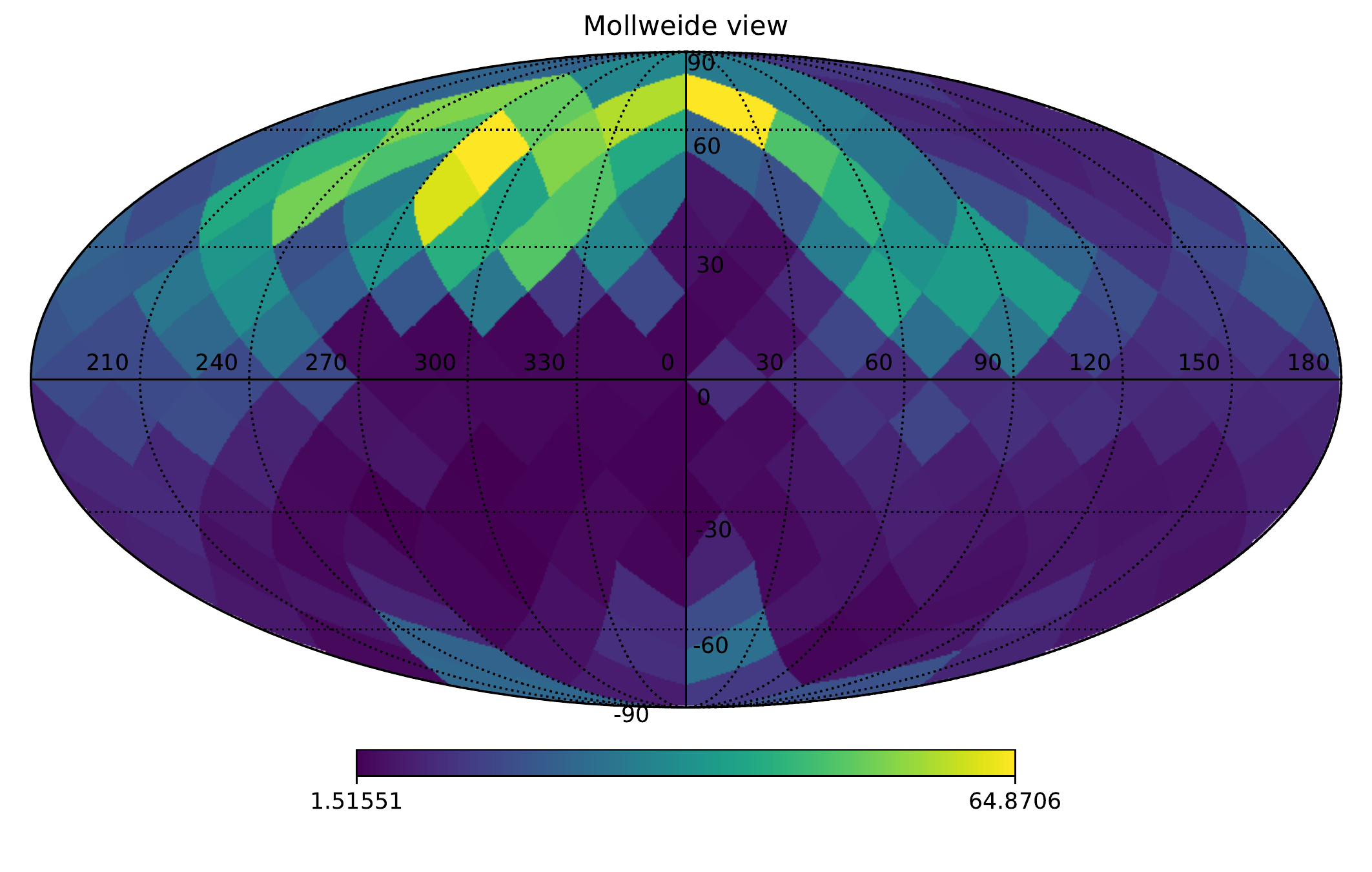}\\
    \includegraphics[width =6cm]{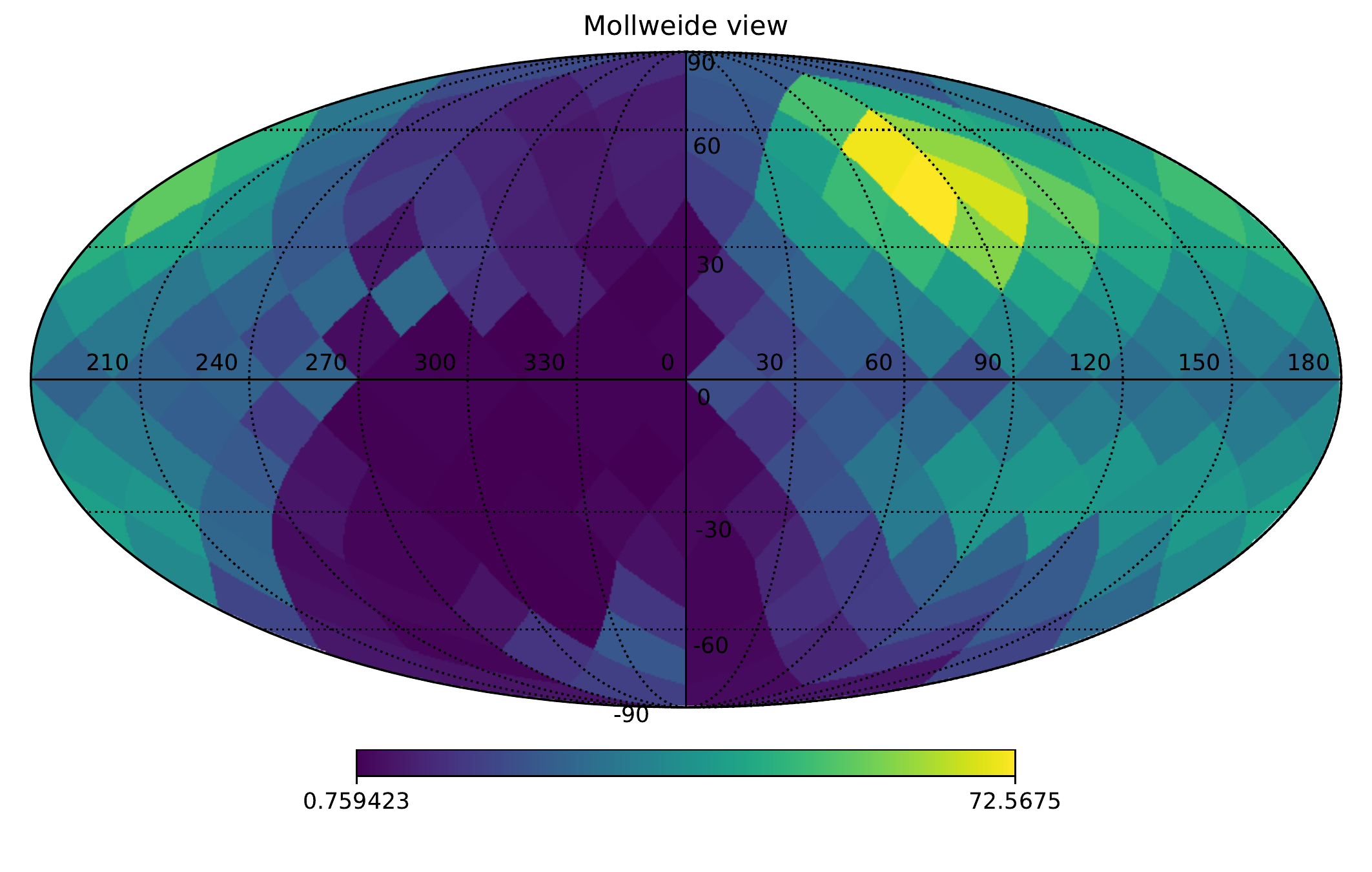} \includegraphics[width = 6cm]{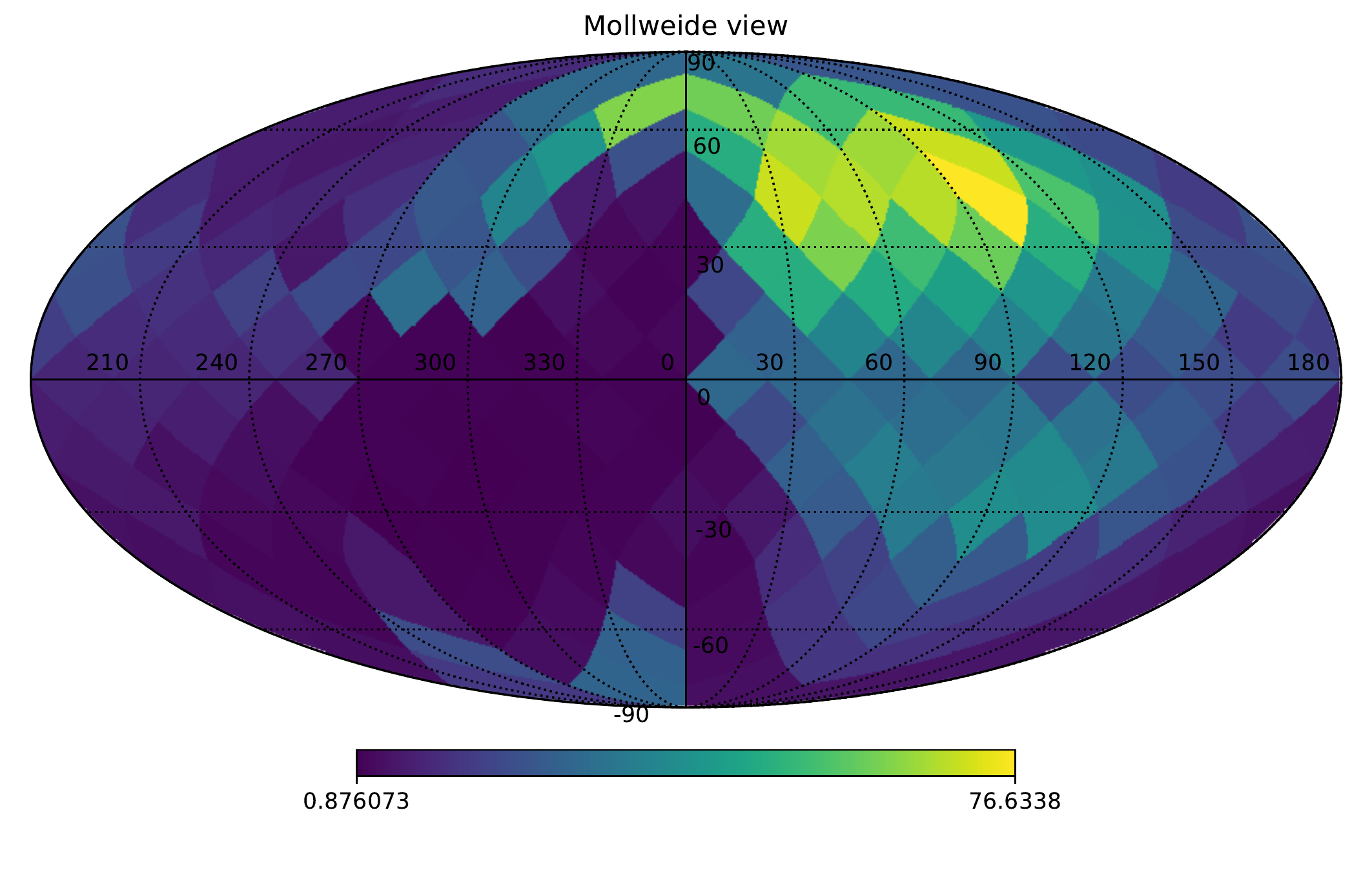}
    \caption{Net effective area for Crab spectrum for each quadrant (A, B, D, C, respectively) as a function of relative $\theta,\phi$ in satellite frame. The top of each plot corresponds to the pointing axis of the satellite. The low effective area regions correspond to lines of sight going through the entire satellite.}
    \label{fig:quad}
\end{figure*}

\subsection{Effective area and source flux}
The sensitivity of CZTI to off-axis sources is a strong function of direction, owing to varying degrees of obscuration by various satellite elements. This has been calculated in detail using the CZTI Mass Model \citep{Mate2021}. Here, we use an older version of the mass model with minor differences. For each direction in the sky, the mass model gives us the $A(E)$, the effective area for each quadrant as a function of energy. \asat\ has a geometric area of $976~\mathrm{cm}^2$ and the all sky median effective area at $180~\mathrm{keV}$ is $190~\mathrm{cm}^2$ \citep{bhalerao2017cadmium}. We ensure that we add up the areas from ``grade 0'' pixels only.

The expected count rate depends on the source spectrum, and its direction in satellite coordinates. Let the source photon flux as a function of energy be $N(E)$. The flux as a function of energy is given by $F(E) = E\cdot N(E)$. Thus, the total expected counts in a quadrant are given by 
\begin{equation}
    h = \int_{E_{\rm{min}}}^{E_{\rm{max}}} A(E) N(E) dE \label{eq:totcounts}
\end{equation}
We define the net effective area for each quadrant by weighting the area at each energy with the source spectrum:
\begin{equation}
A_\mathrm{eff} = \frac{\int_{E_{\rm{min}}}^{E_{\rm{max}}} A(E) F(E) dE}{\int_{E_{\rm{min}}}^{E_{\rm{max}}}F(E) dE}
\end{equation}

Despite the physical proximity of the four quadrants, the effective area in the same direction can be quite different, especially when the collimators of one quadrant casts a shadow on another. We show the all-sky net effective area for all four quadrants for a Crab-like spectrum in Figure~\ref{fig:quad}. It is clear that effective area varies by more than an order of magnitude over the entire sky.

In practice, we assume that the source spectrum is a power law of the form $N(E) = N_0 E^{\Gamma}$, and use Equation~\ref{eq:totcounts} to calculate $N_0$ from the measured values of $h$. We then calculate the total source flux as $F = \int_{E_{\rm{min}}}^{E_{\rm{max}}} A(E) N_0 E^{\Gamma + 1} dE $.

\subsection{Outlier rejection and vetos}\label{sec:veto}
Low quality data with high noise could lead to false or inaccurate measurements. Hence, we define various quality cuts on the data to discard potentially problematic parts of the data. Our pipeline employs five vetos: the first ($\beta$-veto) is applied on all quadrants jointly, while others are applied to individual quadrant data. If only a single quadrant survives the vetos, we discard the entire event.
\begin{enumerate}
  \item Events with $\beta > 60^{\circ}$ have high $\tau$ values, corresponding to very long transitions which are difficult to effectively decouple from the background. Hence, such events are not processed.
  \item Based on our estimates of the minimum detectable transit amplitude ($h_\mathrm{min}$), we conclude that very small values of $h$ are indicative of a failed fit. These are typically caused by noisy backgrounds or low effective area in the direction of the source. Guided by our calculation that $h_\mathrm{min} \approx 0.9$, we veto out any occultation where $|h| < 1$. 
  \item Some data sets are particularly noisy, and the best-fit value of $h$ has a wrong sign: showing an increase in counts at ingress or a decrease in counts at egress. These are discarded by comparing the obtained sign of $h$ with the expected sign.
  \item  The effective area of the detector to the source has a direct correlation with observed source photon count rate from the source. Only allowing events with good effective area ensures we can always expect a reasonable source photon count rate in our data and hence, a more confident fit. We accept only those events where the 4-quadrant effective area is $>10~\mathrm{cm}^2$.
  \item Since the four quadrants are independent, so are their noise properties. If only one quadrant is noisy, data from the others may still prove to be useful. We cannot directly compare the $h$ values of different quadrants owing to differences in their effective areas. We convert the $h$ to fluxes, then calculate the mean and standard deviation of the flux values for all four quadrants. We discard any quadrants where the flux is outside a 1-sigma interval centred on the mean. 
\end{enumerate}

\section{Results}
We apply this method to two bright astrophysical sources: First, we test the methods on the Crab to measure its flux. Second, we select the variable source \cyg\ and measure its flux as a function of time.

\subsection{Crab pulsar and nebula}  

We use a simplified power-law spectral model for the Crab pulsar and nebula, $F(E) = F_0 E^{\Gamma + 1}$ with photon power law index $\Gamma = -2.1$ and normalisation $F_0 = 9.7~\mathrm{keV~cm^{-2}~s^{-1}~keV^{-1}} =  1.5\times10^{-8}~\mathrm{erg~cm^{-2}~s^{-1}~keV^{-1}}$ at 1~keV \citep{2017ApJ...841...56M}. With this model, the Crab flux in the 20--200~keV band is $2.4 \times 10^{-8}~\mathrm{erg~cm^{-2}~s^{-1}}$.
We note that this simplistic treatment does not account for the hard X-ray spectral breaks present in the Crab spectrum \citep[see for instance][]{2013PASJ...65...74K}, but suffices for our purposes. 

We run our codes on on readily available data across five years of CZTI observation, which span about ten thousand orbits. Owing to the low declination of Crab ($\delta = 22\degr$), this results in over 54,000 ingress / egress events in individual quadrants (about 25\% of total possible events). Only 5559 events pass our veto filters.
By treating all quadrants independently, we calculate the 20--200~keV Crab flux to be $(2.40 \pm 0.02) \times 10^{-8}~\mathrm{erg~cm^{-2}~s^{-1}}$. If we instead add another intermediate step by combining $h$ values and effective areas of all quadrants that passed the veto for any given event, and then combine results across all events, we measure the flux as $(2.65 \pm 0.02) \times 10^{-8}~\mathrm{erg~cm^{-2}~s^{-1}}$, where $0.02 \times 10^{-8}~\mathrm{erg~cm^{-2}~s^{-1}}$ in both measurements is the weighted error $\sigma_{\bar{x}}$, using individual error $\sigma_i$  computed as,
\begin{align}
    \sigma_{\bar{x}} = \sqrt{\frac{ 1 }{\sum_{i=1}^n \sigma_i^{-2}}}.\label{we}
\end{align}

\subsection{Higher effective area}
\begin{figure}[ht]
    \centering
    \includegraphics[width = 0.8\linewidth]{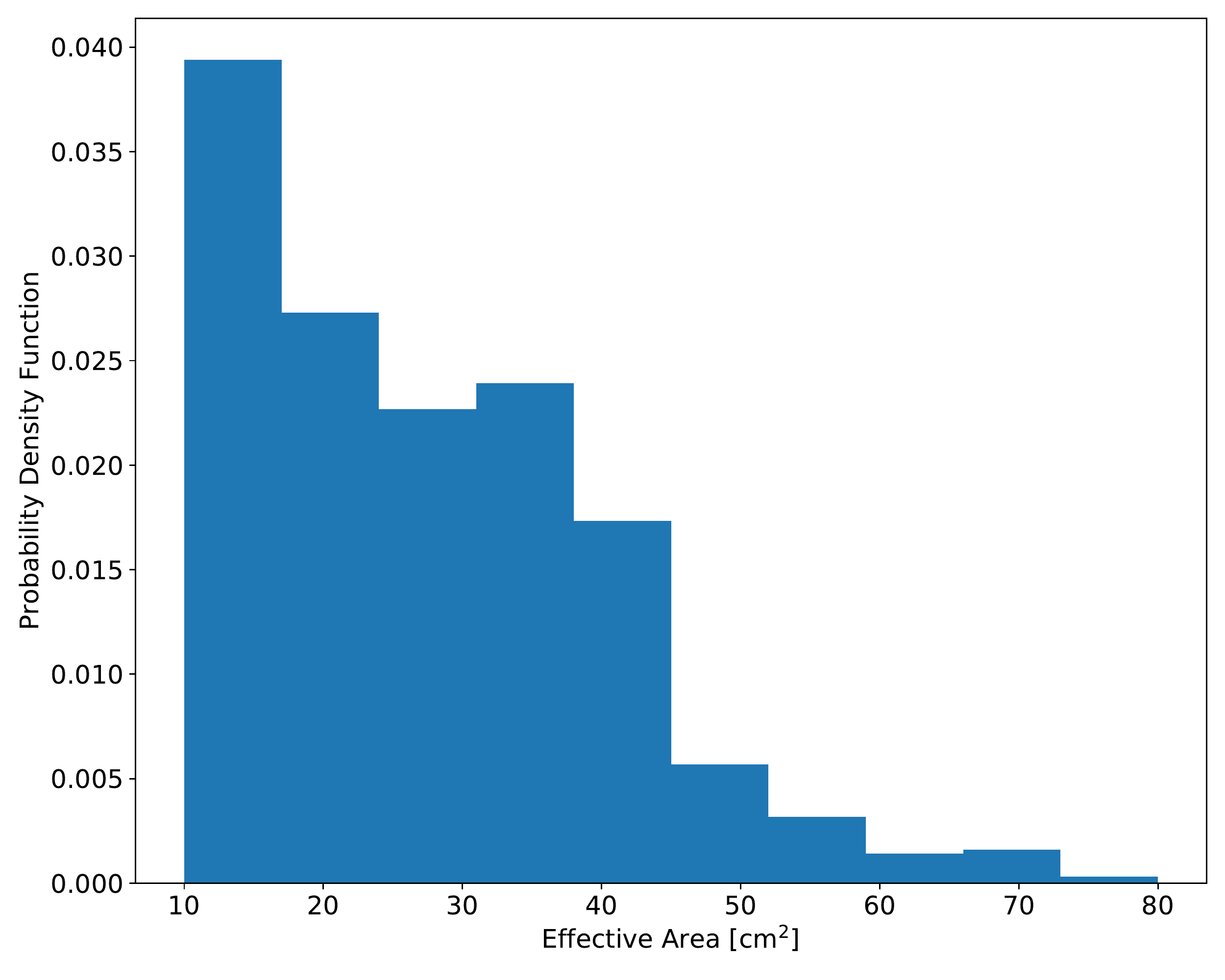}
    \caption{Effective area for the Crab's spectrum across all events.}
    \label{fig:crab_his}
\end{figure}

The amplitude of the measured signal ($h$) is directly related to the effective area in the direction of the source. Through the course of routine CZTI observations, Crab occultations have occurred at a wide variety of angles in the CZTI reference frame, corresponding to a wide range of effective areas. Figure~\ref{fig:crab_his} shows a histogram single-quadrant effective areas calculated from all observable occultations. We then start raising the effective area threshold, by selecting only events with higher effective areas. If we select the top 10\% events with the highest effective area (corresponding to $A_\mathrm{eff} \gtrsim 41~\mathrm{cm}^2$), we measure the Crab flux to be $(2.25 \pm 0.02) \times 10^{-8}~\mathrm{erg~cm^{-2}~s^{-1}}$. Values for other percentile selections along with weighted error $\sigma_{\bar{x}}$  and median error of $\sigma_i$  are given in Table~\ref{tab:1}.

\begin{table}[ht]
    \centering
    \begin{tabular}{|p{1.65cm}|x{1.1cm}|p{0.8cm}|x{1.4cm}|x{1.5cm}|}
      \hline
      \%~selected & Area cut  & Flux &Median error& Weighted error\tabularnewline
      \hline
       & cm$^2$ &        \multicolumn{3}{c|}{$10^{-8}$~erg~cm$^{-2}$~s$^{-1}$}\tabularnewline
      \hline
      All & $>10$ & 2.65 & 1.20&0.02\tabularnewline
      Top 50\% & $\gtrsim 25$ & 2.41 & 0.95&0.01\tabularnewline
      Top 20\% & $\gtrsim 35$ & 2.29 & 0.81&0.01\tabularnewline
      Top 10\% & $\gtrsim 41$ & 2.25 & 0.75&0.02\tabularnewline
      \hline
    \end{tabular}
    \caption{
    Refining Crab flux measurements based on combining all the quadrants by selecting a progressively smaller subsets of events with the highest effective area. The first column gives the selection cut applied above the default veto cut that requires the quadrant effective area to be at least 10~cm$^2$. The second column gives the corresponding effective area cutoff for our sample, in cm$^2$. The third column lists the flux measured from this sample, in units of $10^{-8}$ erg~cm$^{-2}$~s$^{-1}$. Fourth and the fifth columns gives nominal error measured in individual occultation event and weighted error as defined in Eq.~(\ref{we}) in the same unit of $10^{-8}$ erg~cm$^{-2}$~s$^{-1}$. The true flux in this energy range in $2.38\times 10^{-8}$ erg~cm$^{-2}$~s$^{-1}$. }
    \label{tab:1}
\end{table}

\subsection{Higher energy ranges}\label{sec:highenergy}
While these results are consistent with the expected flux, they have large uncertainties. CZTI data have relatively higher electronic noise contribution below 50~keV. The energy range from 50--70~keV is dominated by fluorescence emission from the Tantalum present in the collimators, hence it is difficult to discern any source spectral properties in this range. In order to improve our measurements, we repeat the analysis by restricting ourselves to the 70--200~keV energy range. The expected Crab flux in this range is $1.0 \times 10^{-8}~\mathrm{erg~cm^{-2}~s^{-1}}$. We re-calculate the effective area for each event in this energy range, and select the top 10\% events with the highest effective area. This corresponds to a 70--200~keV effective area cut of 33~cm$^2$, and gives us 622 ingress or egress events. The measured flux from this sample is $(1.38~\pm~0.02)~\times~10^{-8}~\mathrm{erg~cm^{-2}~s^{-1}}$ and median individual error of $1.17 \times 10^{-8}~\mathrm{erg~cm^{-2}~s^{-1}}$.

\subsection{\cyg}\label{sec:cyg}
\begin{figure*}[ht]
    \centering
    \includegraphics[width=17cm]{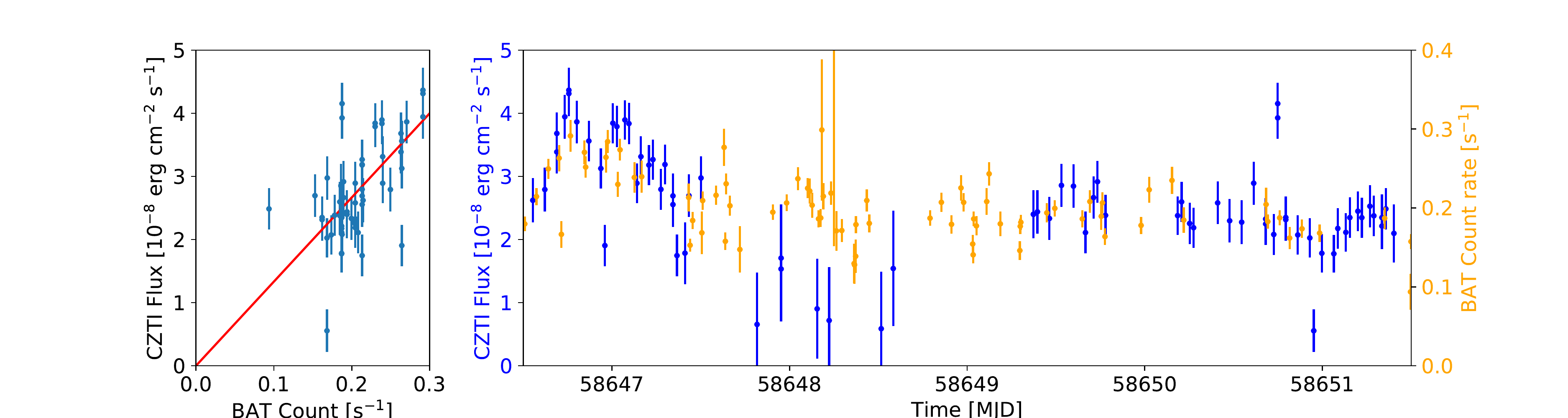}
    \caption{\textit{Left:} Shows direct comparison between the CZTI fluxes with BAT count rates. Here we have assumed zero errors for BAT values. \textit{Right:} Comparing \cyg\ measurements from {\em Swift}-BAT and CZTI. Blue filled circles with error bars denote flux measurements by CZTI using EOT. Orange crosses are BAT count rates. BAT rates are scaled to CZTI fluxes as discussed in \S\ref{sec:cyg}. }
    \label{fig:cyg_S_A}
\end{figure*}

We then use our EOT pipeline to measure the flux of \cyg. To limit ourselves to data with high effective areas, we select two on-axis observations of \cyg\ conducted in June 2019, and analyse the 70 ingress or egress events that occurred in this duration. We note that despite the source being on-axis, we do not utilise any coded aperture mask data processing for measuring fluxes. \cyg\ is highly variable, and expected to be visible to us only in the bright hard state. Following \citet{2020ApJ...896..101L}, we model the ``pure hard'' state \cyg\ spectrum as a power-law with a photon index $\Gamma = -1.7$. For comparison of our measurements, we obtain \cyg\ data from the Swift/BAT Hard X-ray Transient Monitor\footnote{ \url{https://swift.gsfc.nasa.gov/results/transients/BAT_current.html\#anchor-CygX-1}} \citep{2013ApJS..209...14K}.
In Figure~\ref{fig:cyg_S_A}, we plot our flux measurements alongside {\em Swift}-BAT count rates. 
\citet{2020ApJ...896..101L} show that a BAT count rate of 0.2 counts~s$^{-1}$ corresponds to an {\em Integral}-ISGRI flux of $\sim 2\times 10^{-8}~\mathrm{erg~cm^{-2}~s^{-1}}$ in the 22--100~keV band. Extrapolating this value to our 20-200~keV band, we expect the corresponding CZTI flux to be $2\times 10^{-8}~\mathrm{erg~cm^{-2}~s^{-1}}$ from independent and individual qaudrants. 
As seen in Figure~\ref{fig:cyg_S_A}, there is very good correspondence between our individual fluxes and BAT count rates, further validating our method.

\section{Discussion}
The primary purpose of CZTI is the Hard X-ray imaging and spectroscopy of on-axis sources by using a coded aperture mask. The field of view is limited to a full-width at half maximum of $4.6\degr \times 4.6\degr$ by using collimators.
Despite this, the reduced opacity of the instrument and satellite structure at energies above $\sim$100~keV makes it an effective all-sky monitor, capable of detecting transients at large off-axis angles. 

In this work, we have extended this capability to the study of persistent sources at large off-axis angles. We demonstrate a proof-of-concept by measuring the flux of the Crab nebula and pulsar, and measuring the variability of \cyg. To the best of our knowledge, this is the first demonstration of measuring source fluxes by using a limited-field collimated instrument. We conclude that the relatively low off-axis effective area limits the applicability of the Earth Occultation Technique to the brightest sources, with fluxes $\gtrsim 1$Crab. This method will be highly effective with future open-detector missions like {\em Daksha} which have a high effective area over the entire sky.

\section*{Acknowledgements}
We thank Vedant Shenoy and Akash Anumarlapudi for their assistance in data analysis. 

CZT--Imager is built by a consortium of Institutes across India. The Tata Institute of Fundamental Research, Mumbai, led the effort with instrument design and development. Vikram Sarabhai Space Centre, Thiruvananthapuram provided the electronic design, assembly and testing. ISRO Satellite Centre (ISAC), Bengaluru provided the mechanical design, quality consultation and project management. The Inter University Centre for Astronomy and Astrophysics (IUCAA), Pune did the Coded Mask design, instrument calibration, and Payload Operation Centre. Space Application Centre (SAC) at Ahmedabad provided the analysis software. Physical Research Laboratory (PRL) Ahmedabad, provided the polarisation detection algorithm and ground calibration. A vast number of industries participated in the fabrication and the University sector pitched in by participating in the test and evaluation of the payload.

The Indian Space Research Organisation funded, managed and facilitated the project.

This work utilised various software including Python, AstroPy \citep{astropy}, NumPy \citep{numpy}, and Matplotlib \citep{matplotlib}.

\vspace{-1em}

\balance
\bibliography{reference}

\begin{thebibliography}{}
\expandafter\ifx\csname natexlab\endcsname\relax\def\natexlab#1{#1}\fi

\bibitem[{Anumarlapudi {$et~al$.}(2020)Anumarlapudi, Bhalerao, Tendulkar, \&
  Balasubramanian}]{anumarlapudi2020prompt}
Anumarlapudi, A., Bhalerao, V., Tendulkar, S.~P., \& Balasubramanian, A. 2020,
  The Astrophysical Journal, 888, 40

\bibitem[{{Anusree} {$et~al$.}(2021){Anusree}, {Bhattacharya}, {Rao},
  {Vadawale}, {Bhalerao}, \& {Vibhute}}]{2021arXiv210108650A}
{Anusree}, K.~G., {Bhattacharya}, D., {Rao}, A.~R., {$et~al$.} 2021, arXiv
  e-prints, arXiv:2101.08650

\bibitem[{Bhalerao {$et~al$.}(2017)Bhalerao, Bhattacharya, Vibhute, Pawar, Rao,
  Hingar, Khanna, Kutty, Malkar, Patil, {$et~al$.}}]{bhalerao2017cadmium}
Bhalerao, V., Bhattacharya, D., Vibhute, A., {$et~al$.} 2017, Journal of
  Astrophysics and Astronomy, 38, 31

\bibitem[{Harmon {$et~al$.}(2002)Harmon, Fishman, Wilson, Paciesas, Zhang,
  Finger, Koshut, McCollough, Robinson, \& Rubin}]{harmon2002burst}
Harmon, B., Fishman, G., Wilson, C., {$et~al$.} 2002, The Astrophysical Journal
  Supplement Series, 138, 149

\bibitem[{Hunter(2007)}]{matplotlib}
Hunter, J.~D. 2007, Computing in Science {\&} Engineering, 9, 90

\bibitem[{{Kapahi} {$et~al$.}(1973){Kapahi}, {Joshi}, {Subrahmanya}, \&
  {Gopal-Krishna}}]{1973AJ.....78..673K}
{Kapahi}, V.~K., {Joshi}, M.~N., {Subrahmanya}, C.~R., \& {Gopal-Krishna}.
  1973, The Astronomical Journal, 78, 673

\bibitem[{{Kouzu} {$et~al$.}(2013){Kouzu}, {Tashiro}, {Terada}, {Yamada},
  {Bamba}, {Enoto}, {Mori}, {Fukazawa}, \& {Makishima}}]{2013PASJ...65...74K}
{Kouzu}, T., {Tashiro}, M.~S., {Terada}, Y., {$et~al$.} 2013, Publications of
  the Astronomical Society of Japan, 65, 74

\bibitem[{{Krimm} {$et~al$.}(2013){Krimm}, {Holland}, {Corbet}, {Pearlman},
  {Romano}, {Kennea}, {Bloom}, {Barthelmy}, {Baumgartner}, {Cummings},
  {Gehrels}, {Lien}, {Markwardt}, {Palmer}, {Sakamoto}, {Stamatikos}, \&
  {Ukwatta}}]{2013ApJS..209...14K}
{Krimm}, H.~A., {Holland}, S.~T., {Corbet}, R.~H.~D., {$et~al$.} 2013, The
  Astrophysical Journal Supplement Series, 209, 14

\bibitem[{{Lubi{\'n}ski} {$et~al$.}(2020){Lubi{\'n}ski}, {Filothodoros},
  {Zdziarski}, \& {Pooley}}]{2020ApJ...896..101L}
{Lubi{\'n}ski}, P., {Filothodoros}, A., {Zdziarski}, A.~A., \& {Pooley}, G.
  2020, The Astrophysical Journal, 896, 101

\bibitem[{{Madsen} {$et~al$.}(2017){Madsen}, {Forster}, {Grefenstette},
  {Harrison}, \& {Stern}}]{2017ApJ...841...56M}
{Madsen}, K.~K., {Forster}, K., {Grefenstette}, B.~W., {Harrison}, F.~A., \&
  {Stern}, D. 2017, The Astrophysical Journal, 841, 56

\bibitem[{{Mate} {$et~al$.}(2021){Mate}, {Chattopadhyay}, {Bhalerao}, {Aarthy},
  {Balasubramanian}, {Bhattacharya}, {Gupta}, {Kutty}, {Mithun}, {Palit},
  {Rao}, {Saraogi}, {Vadawale}, \& {Vibhute}}]{Mate2021}
{Mate}, S., {Chattopadhyay}, T., {Bhalerao}, V., {$et~al$.} 2021, Journal of
  Astrophysics and Astronomy accepted for publication

\bibitem[{Robitaille {$et~al$.}(2013)Robitaille, Tollerud, Greenfield,
  Droettboom, Bray, Aldcroft, Davis, Ginsburg, Price-Whelan, Kerzendorf,
  Conley, Crighton, Barbary, Muna, Ferguson, Grollier, Parikh, Nair,
  G{\"{u}}nther, Deil, Woillez, Conseil, Kramer, Turner, Singer, Fox, Weaver,
  Zabalza, Edwards, {Azalee Bostroem}, Burke, Casey, Crawford, Dencheva, Ely,
  Jenness, Labrie, Lim, Pierfederici, Pontzen, Ptak, Refsdal, Servillat, \&
  Streicher}]{astropy}
Robitaille, T.~P., Tollerud, E.~J., Greenfield, P., {$et~al$.} 2013, Astronomy
  {\&} Astrophysics, 558, A33

\bibitem[{{Sharma} {$et~al$.}(2020){Sharma}, {Marathe}, {Bhalerao}, {Shenoy},
  {Waratkar}, {Nadella}, {Page}, {Hebbar}, {Vibhute}, {Bhattacharya}, {Rao}, \&
  {Vadawale}}]{sharma2020search}
{Sharma}, Y., {Marathe}, A., {Bhalerao}, V., {$et~al$.} 2020, arXiv e-prints,
  arXiv:2011.07067

\bibitem[{Singh {$et~al$.}(2014)Singh, Tandon, Agrawal, Antia, Manchanda,
  Yadav, Seetha, Ramadevi, Rao, Bhattacharya, {$et~al$.}}]{singh2014astrosat}
Singh, K.~P., Tandon, S., Agrawal, P., {$et~al$.} 2014, in Space Telescopes and
  Instrumentation 2014: Ultraviolet to Gamma Ray, Vol. 9144, International
  Society for Optics and Photonics, 91441S

\bibitem[{van~der Walt {$et~al$.}(2011)van~der Walt, Colbert, \&
  Varoquaux}]{numpy}
van~der Walt, S., Colbert, S.~C., \& Varoquaux, G. 2011, Computing in Science
  {\&} Engineering, 13, 22

\bibitem[{Wilson-Hodge {$et~al$.}(2012)Wilson-Hodge, Case, Cherry, Rodi,
  Camero-Arranz, Jenke, Chaplin, Beklen, Finger, Bhat,
  {$et~al$.}}]{wilson2012three}
Wilson-Hodge, C.~A., Case, G.~L., Cherry, M.~L., {$et~al$.} 2012, The
  Astrophysical Journal Supplement Series, 201, 33

\bibitem[{Zhang(2013)}]{zhang2013searching}
Zhang, Y. 2013, PhD thesis, LSU Doctoral Dissertations

\end{thebibliography}

\end{document}